\documentclass[prc,aps,floatfix,nofootinbib,preprint]{revtex4-1} 
\usepackage{xcolor}
\usepackage{graphicx} 
\usepackage{amssymb} 
\usepackage{amsmath} 
\usepackage{amsfonts} 
\def\lb{\langle} 
\def\rb{\rangle} 
\def\be{\begin{equation}} 
\def\ee{\end{equation}} 
\def\eps{\varepsilon}
\def\unf{$^{235}$U(n,f)}
\begin{document} 
\title{Fluctuations in the $^{235}$U(n,f) cross section
} 
\author{G.F.~Bertsch } 
\affiliation{
Department of Physics and Institute of Nuclear Theory, 
Box 351560\\ University of Washington, Seattle, Washington 98915, USA} 
\author{David Brown} 
\affiliation{National Nuclear Data Center,
Brookhaven National laboratory, Upton NY, USA}
\author{E.D.~Davis}
\affiliation{
Department of Physics, North Carolina State University, Raleigh, North Carolina 27695-8202, USA}
\affiliation{
Triangle Universities Nuclear Laboratory, Durham, North Carolina 27708-0308, USA}
 
\begin{abstract} 
 
We examine the autocorrelation function of the $^{235}$U(n,f) reaction
with a view to quantify the presence of 
intermediate structure in the cross section.  Fluctuations due to
compound nucleus resonances on the eV energy scale are clearly 
visible up to $\sim100$ keV neutron energies.  Structure on the one-keV energy
scale is not present as a systematic feature of the correlation function,
although it is present in the data covering the region around 20
keV. \\
{\tt autocorr.12.tex}
\
\end{abstract}

\maketitle 
 
\section{Introduction} 

  The fluctuations in reaction cross sections convey important information
about the reaction dynamics.  Low-energy reactions on heavy nuclei are
typically described by the Hauser-Feshbach extension of Bohr's compound
nucleus model, and contain as parameters the density of
the compound nucleus resonances and the effective number of channels
participating in the reaction.  When the properties of the resonances
cannot be individually measured, one relies on a fluctuation analysis
such as we use below to gain information about their properties.
Understanding the neutron-induced fission reaction is even more
challenging because there is no reason to believe that a simple
compound nucleus picture is adequate to deal with the large-amplitude
shape changes the excited nucleus undergoes.  This gives a strong motivation
to characterize as quantitatively as possible the fluctuations
associated with the reaction, in order to better understand the
reaction dynamics.

Part of the complexity of the reaction is due to the presence of
multiple fission barriers.  At energies below the barriers, the cross
section can fluctuate due resonant states located between the barriers. 
Above the barriers, the situation remains unclear. 
There is evidence for structure on the keV energy scale
from experiments carried out in the 1970's 
on the $^{235}$U(n,f) reaction with neutron energies in the range
1--100 keV.  The compound nucleus energies exceed the barrier top by
$\sim 1$ MeV, so below-barrier resonances could not be the cause.
Many experimental measurements have been made encompassing 
that energy range, as detailed in Appendix A.
Three of them have the resolution and documentation to make a
case for the presence of fluctuations on a keV energy scale.
Their measured cross sections between 10 and 25 keV are shown in Fig. 1
All three show a clear peak at 22 keV, having
a width about 1 keV.  There may be correlated peaking at lower
energies as well.  The abstract of one of the papers (Ref.~\cite{pe74}) states:
``The previously reported intermediate structure in the fission
cross section in the keV region is confirmed by the results of
this work."
\begin{figure}[tb] 
\begin{center} 
\includegraphics[width=0.5\columnwidth]{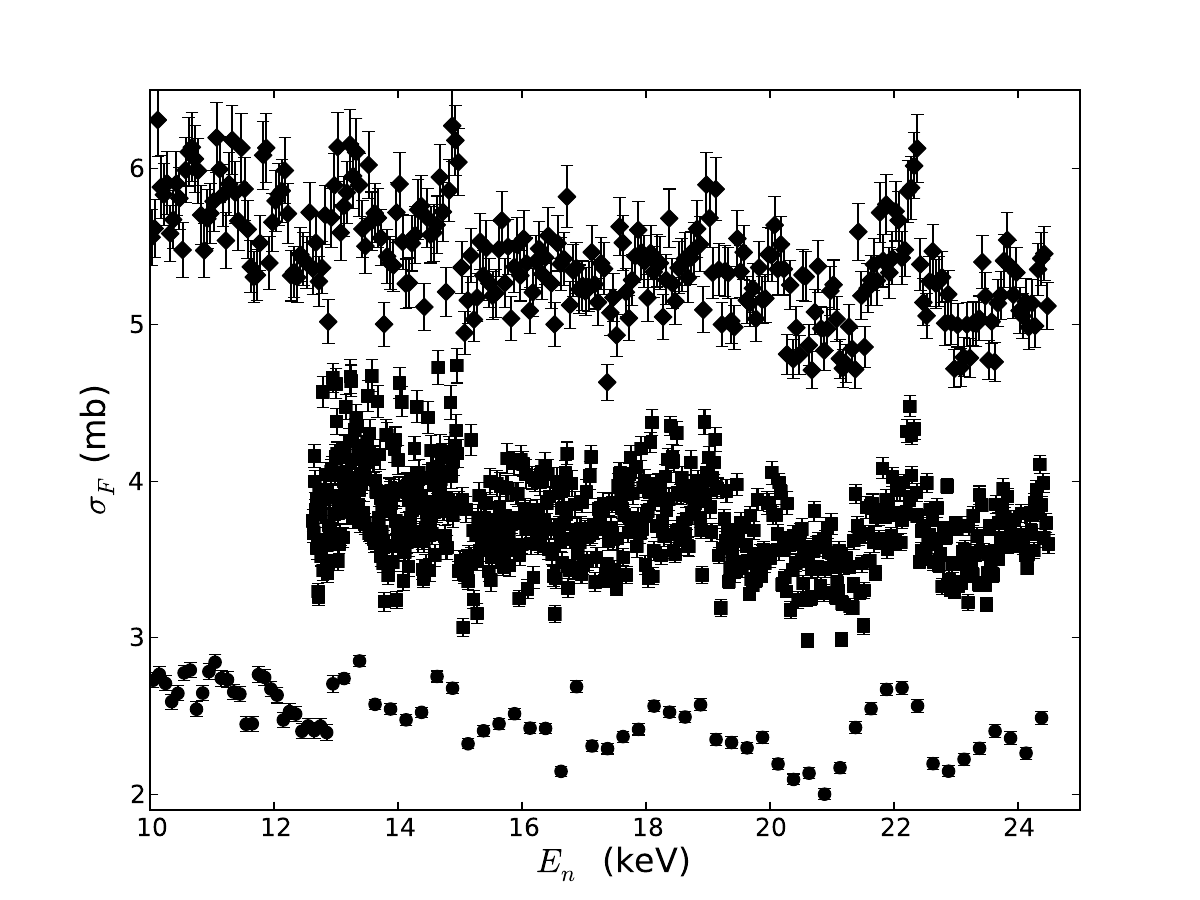}
\caption{Experimental fission cross sections in the range 10 - 25 keV.
Circles: from Ref. \cite{mo78}; squares: from \cite{bo71}; diamonds:
from \cite{pe74}.
The latter two data sets have been shifted upward
for clarity in the plotting of the figure.
}
\label{allb} 
\end{center} 
\end{figure} 
Nevertheless, such structures have not be 
documented as a general feature of cross
sections at energies above the barrier.  The aim of this work is
to analyze the cross section fluctuations in a systematic way to
see if quantitative information about them can be extracted.

Our analysis tool is the autocorrelation function $R$ defined
as~\cite{AGH64,Ha64}
\be
\label{R}
R(\eps)= \left\langle  
{(\sigma(E_+)- \bar \sigma(E_+)(\sigma(E_-)- \bar \sigma(E_-)
\over \bar\sigma(E_+) \bar \sigma(E_-)}\right\rangle
\ee 
where $E_\pm = E \pm \eps/2$. The angle   
brackets denote an average over the energy $E$ and 
$\bar \sigma$ is an energy-dependent averaged cross section, with the details
to be specified later.
To keep the number of entrance channels to a minimum, we limit the
analysis to neutron energies below 100 keV, which is sufficient to span
the structure of interest at 22 keV.

The organization of this article 
is as follows.  In Section II, we
review the interpretation of the autocorrelation function and its
parameterization.  In Section III, we confirm the expected behavior
of the autocorrelation function in the isolated-resonance region at the
lowest energies.   The data in the
higher energy region of unresolved resonances is analyzed in Sect. IV.
Section V summarizes the two main conclusions of our analysis.  The first
conclusion is 
that the  eV-scale correlations due to compound
nucleus resonances are present and affect the correlation function far
beyond the isolated resonance region.   Second, only a limit can be placed
on any systematic correlation structure at the one-keV energy scale.  Thus the
peaking seen in Fig. 1 is isolated feature of the energy-dependent cross section.
We argue for a campaign of new measurements 
to pinpoint the origin of the observed structure and to see if it occurs
in above-barrier fission of other nuclei.
\section{The autocorrelation function}
Before discussing practical details of calculating and interpreting
cross section fluctuations, we review the analytic statistical theory of
the autocorrelation function $C(\eps)$. It is defined 
\be \label{thC}
C(\eps ) = 
\overline{\left(\sigma(E_+)- \overline{ \sigma} \right)\left(\sigma(E_-)- \overline{\sigma} \right)}\left/ \overline{\sigma}^2
\right..
\ee
Here the overline denotes an average over an ensemble of $S$-matrix
elements that enter the cross sections, Eq. (3) below.
The ensemble is
generated from a 
statistical  model of compound nucleus resonances. 
The ensemble averages are calculated at fixed energy, but by construction they 
do not depend on that energy.  It is implicitly assumed that the energy
averaging in $R(\eps)$ is equivalent
to the ensemble averaging in $C(\eps)$.

The first step in calculating $C(\eps )$ is to express the cross sections in
terms of their $S$-matrix elements,
\be  \label{eq:cs}
  \sigma = \frac{\pi}{k^2_n}  \sum_{i,c} g^{J} |S_{ic} |^2.
\ee
Here $i$ denotes the quantum numbers specifying an incident (s-wave) neutron channel of
angular momentum $J$ and parity $\pi$, $c$ denotes the quantum numbers of an exit fission 
channel (of the same $J^\pi$), and $g^{J}$ is the usual spin statistical
factor.  $C(\eps )$ can then be 
written as the incoherent superposition
\be \label{eq:cwc}
 C(\epsilon)  = \sum_{i, c, c^{\prime}} w_{ic} w_{ic'} C_{ic c^{\prime}} (\epsilon),
\ee 
of the autocorrelation functions for each channel $i$,  
\be
 C_{ic c^{\prime}}(\eps ) =  \frac{\overline{\,  |S_{ic}(E+\epsilon/2)|^2 |S_{ic^{\prime}}(E-\epsilon/2)|^2 } - 
                                                   \overline{ |S_{ic}|^2 }\,\, \overline{ |S_{ic^{\prime}}|^2 }  }{ 
                                                    \overline{ |S_{ic}|^2 }\,\, \overline{ |S_{ic^{\prime}}|^2 }  }
\ee
together with the weighting factors
\be
  w_{ic}  =  \frac{g^{J} \,  \overline{ |S_{ic} |^2}  }
{\sum\limits_{i,c^{\prime}}  
                                                                              g^{J} \overline{ | S_{ic^{\prime}} |^2 } }  .
\ee
The correlations of interest are determined  by the $C_{icc'}$, but their
amplitude in $C(\eps)$ depends on the number of fission channels and other information
carried by the weights.

We will see later that the 
$\eps$- and $w_{ic}$-dependence are easy to disentangle in the isolated
resonance regime, as well as the regime with strongly overlapping
resonances.  
\subsection{The statistical Breit-Wigner model}\label{sc:aap}
Our derivation of $ C_{ic c^{\prime}}(\eps )$  proceeds by modeling the $S$-matrix by
a sum of Breit-Wigner resonances,
\be \label{eq:sres}
S_{ab}(E) = \delta_{ab} -i \sum_k{ \gamma_{ak} \gamma_{kb} \over E - e_k}. 
\ee
The poles are at complex energies $e_k = E_k - i \Gamma_k/2$, 
where the widths $\Gamma_k$ are related to the real-valued partial width amplitudes $\gamma_{ck}$ 
by $\Gamma_{k} = \sum_c\gamma_{ck}^2$. 
Evaluation of Eq. (5) requires
assumptions about the distribution and correlations of $\gamma_{ck}$ and
$E_k$.  Here we are guided by the empirical success of the Gaussian Orthogonal Ensemble (GOE)
of Hamiltonian matrices.  According to the GOE model, partial width amplitudes $\gamma_{ak}$ are distributed for 
different resonances $k$ as a Gaussian random variable of zero mean; the corresponding
variance depends on the choice of channel $a$ alone. Furthermore,
partial width amplitudes relating to different open channels $a$ are completely 
uncorrelated. As a consequence, for inelastic processes ($a\not=b$), the ensemble
average of $S_{ab}$ in Eq.~(\ref{eq:sres}) yields $\overline{S_{ab}}=0$.

Our treatment of the $e_k$ in the energy denominator deviates from a strict application
of GOE level correlation statistics.
Instead of using Dyson's celebrated result for the two-level cluster 
function $Y_2$~\cite{Dy62}, we follow Ref.~\cite{er16} and adopt
the simplified but effective parameterization 
\be \label{eq:Y2app}
 Y_2(\Delta E/\bar D) \approx \frac{1}{1+(\pi \Delta E/\bar D)^2}
\ee
where $\Delta E$ is the difference of two resonance energies of the same spin and parity, and 
$\bar D$ is the average level spacing for that spin and parity.  The
imaginary part of $e_k$, namely $\Gamma_k/2$ is assumed to be constant.
Neglect of fluctuations in the \emph{total} widths $\Gamma_k$
will be a source of inaccuracy if there are only a few open fission channels. (We return to this point below.)

Note that the analysis of the ensemble is at a fixed energy
and thus not able to deal with secular variations of the parameters
with respect to energy.  Thus the theory does not address effects related
to penetrability factors in the 
amplitudes $\gamma_{ak}$ or to increase of level densities with excitation
energy.
 
\subsection{Analytic approximations for  $C_{icc^{\prime}}(\eps )$}\label{sc:AAfC}
Compact integral representations of the averages 
in $C_{icc^{\prime}}(\eps)$ can be derived using the method presented in Ref.~\cite{er16}.
Our replacement of the different total widths $\Gamma_k$ 
by a single total width $\Gamma$ permits us to
infer explicit expressions which capture many of the
essential features of $C_{icc^{\prime}}(\eps)$. In terms of the ratio $x=\bar D/\pi \Gamma$, 
our full result for $C_{icc^{\prime}}(\eps)$ reads
 \begin{equation}\label{eq2}
 \begin{split}
 C_{icc^{\prime}}(\eps) \approx \;
            & 3(1+2\delta_{c,c^{\prime}})  \frac{x\Gamma^2}{\eps^2+\Gamma^2}  \\
            & - \left(\frac{x}{x+1} \right) \frac{\Gamma^2}{\eps^2+\Gamma^2}
                -\frac{x(x+1)\Gamma^2}{\eps^2+ (x+1)^2\Gamma^2} 
                -  \left(\frac{x}{x+1} \right) \frac{\Gamma^2}{\eps^2+ (x+1)^2\Gamma^2} \\
            & + \delta_{c,c^{\prime}}    \frac{\Gamma^2}{\eps^2+\Gamma^2} .
\end{split}
 \end{equation}
In the limit $\eps=0$ and $c=c^{\prime}$, Eq.~(\ref{eq2}) reduces  to the second 
line of Eq.~(B12) in Ref.~\cite{er16}, 
when allowance is made for the different grouping of terms, the identification of
$\Gamma$ with the Weisskopf estimate $\Gamma_W$ for the correlation width, 
and a typographical error~\cite{Typo}.

The three lines in Eq.~(\ref{eq2}) correspond to three physically distinct contributions to $C_{icc^{\prime}}(\eps)$: on
the first line,
a resonance self-correlation piece, dominant when $\bar D \gg \Gamma$ (isolated resonance regime); 
on the second line, negative terms
arising from level repulsion correlations between pairs of distinct resonances, 
and; finally, on the third line, a cross section auto-correlation function of the kind
derived by Ericson~\cite{er60}, dominant when $\bar D\ll \Gamma$, i.e., in the strongly overlapping resonance regime.

In the limit of large or small $x$, 
Eq.~(\ref{eq2}) depends on $\eps$ only by an overall factor 
$\Gamma^2/(\eps^2+\Gamma^2)$.  In Appendix \ref{simple_formula}  we show that the factorization
can be extended over the entire range of $x$ with only a slight
degradation of accuracy with the formula
\be  \label{eq:Cga}
C_{icc^{\prime}}(\eps) \approx C_{icc'}(0) \frac{\Gamma^2}{\eps^2+\Gamma^2} ,
\ee
where
\be
\label{simple_f}
C_{icc'}(0) \approx\Bigl[ (1+ 6 x)\delta_{cc^{\prime}} + (3x-1)\Theta(3x-1) \Bigr]
\ee
and $\Theta(x)$ is the Heaviside step function. 

With the above approximations, it is easy to carry out the sum
of $c$ and $c'$ in Eq. (\ref{eq:cwc}).  $C(\eps)$ has the form
\be
\label{eq:Csummary}
C(\eps) = C(0) {\Gamma^2 \over \eps^2 + \Gamma^2}
\ee
where  
\be \label{eq:C0}
 C(0)\approx (1+6 x)\sum_{i,c} w_{ic}^2 + (3x -1)\Theta(3x-1)
\sum_{i,c,c'} w_{ic} w_{ic'} ,
\ee
Note that the
two sums depend on the number of significant entrance and exit
channels. The sum $\sum_{i,c} w_{ic}^2$ in the first term of Eq.
(\ref{eq:C0}) appears in the theory for strongly overlapping resonances
~\cite{Er63,Ha64,BL65,Gi67} as the damping factor 
$N^{-1}$. 
However, Eq. (\ref{eq:C0}) applies to a broader range of conditions
than the formulas derived in these publications.

\subsection{Extracting $R$ from experimental data}

Several compromises must be made to apply Eq. (1) and interpret the results.
For the data treated here in detail, the experimental cross sections are
provided as average cross sections on a mesh of energies with mesh
spacing $\Delta E$
covering some range of energy $[E_0 - B/2,E_0 + B/2]$. The cross
sections are given as a list $\sigma_n$ where $n$ specifies the energy
$E_n = E_0 - B/2 + n \Delta E$.  The ratio in Eq. (1) is computed
for each pair of cross sections in the list, and a running sum
$\Sigma(|n-n'|)$ is kept for each energy difference $E_n-E_{n'}$.
The computed autocorrelation function is $R(\eps) = \lb \Sigma(m)\rb$
where the angular brackets denote the average for that bin.

In the analysis presented below, we define a local average cross section
$\langle \sigma(E) \rangle_l$ by making a linear fit 
\be
\langle \sigma(E) \rangle_l = \sigma_0 + b (E-E_0)
\ee
to the data in the interval.
We have also analyzed some of the data with a quadratic fit and found
that the extracted correlation is hardly changed.  One can develop
a semi-analytic justification based on the assumption that the actual variation of the local
average cross section is due to the $s$-wave penetrability factor,
giving $\langle \sigma (E)\rangle_l \sim E^{-1/2}$.  With our bandwidths $B$, which are such that
$B/E_0 < 1 $, a linear fit produces an error of less than 1\% in
$R(0)$.

The analysis will show a peak at $\epsilon = 0$ which may or may not
extend to other bins.  The experimental statistics are not good enough
to test the actual shape of peak, but we can extract $R(0)$ and
some measure of the peak width within the experimental uncertainty 
limits of the data.  We shall extract an ``experimental" $\Gamma$
as the value of $\eps$ satisfying  $R(\eps) = R(0)/2$.

Equation (\ref{eq:Csummary}) does not take into account the finite energy resolution
of an experi{\-}mental measurement.  Typically, cross sections are reported
as averages over energy bins $\Delta E$.  The effect on $R(\eps)$ is
analyzed in Appendix C.  
In the limit $\eps \gg \Delta E$ the peak occurs only in the first bin,
and its height $R_{\Delta E}$ is
\be
\label{eqRDE}
R_{\Delta E} \approx R(0) 
 {\pi\Gamma\over \Delta E}.
\ee
Equation (\ref{eqRDE}) applies to the data sets we consider in section IV as these involve multi-keV 
neutron energies, for which the experimental resolution is much broader
than the widths of the compound-nucleus
states.

\section{Resolved resonance region}

\begin{figure}[tb] 
\begin{center} 
\includegraphics[width=0.5\columnwidth]{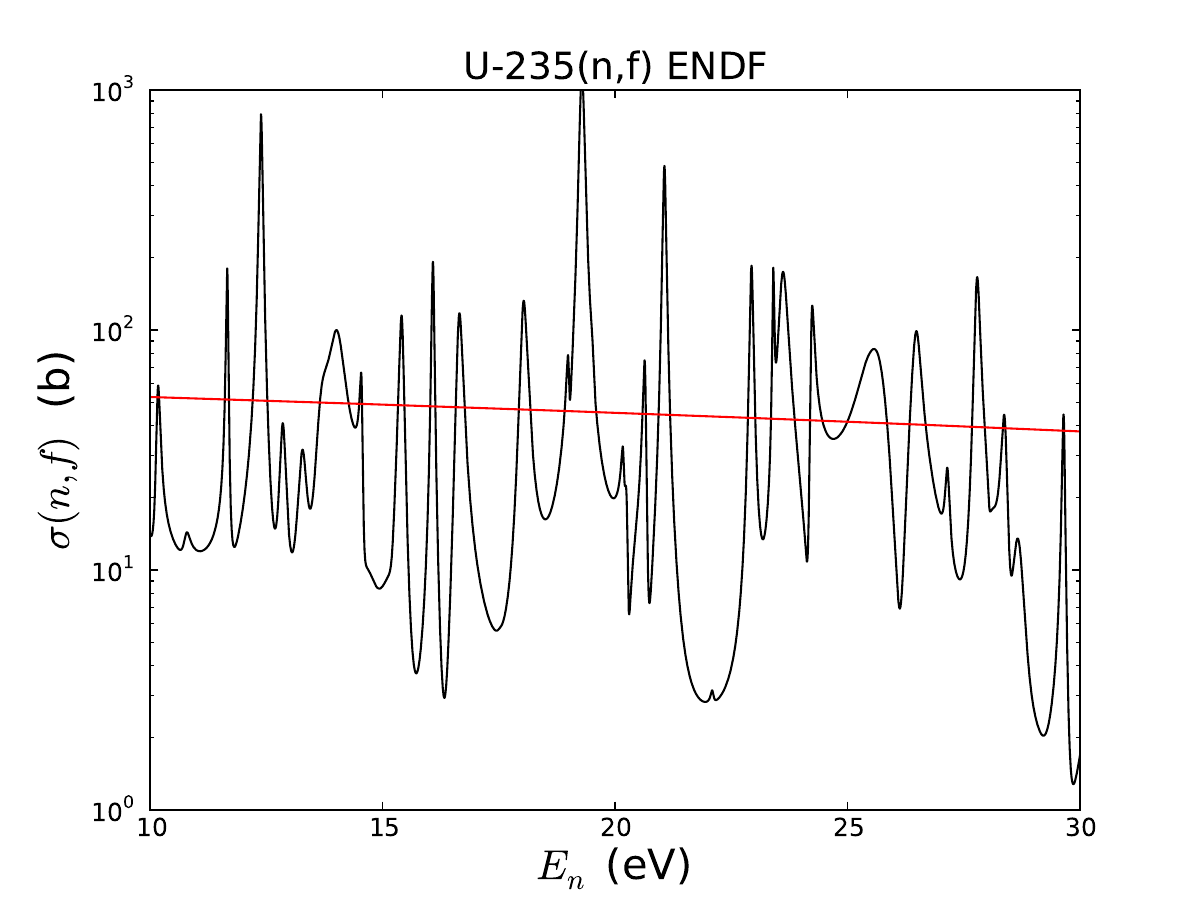}
\caption{$^{235}$U(n,f) cross section from the ENDF/VIII.0 database. The red line
shows the fitted average cross section, 
$\bar \sigma= 45.2 -0.74 (E_n - 20 \rm{~eV}) $ b.}
\label{sigma-endf}
\end{center} 
\end{figure}

\begin{figure}[tb] 
\begin{center} 
\includegraphics[width=0.5\columnwidth]{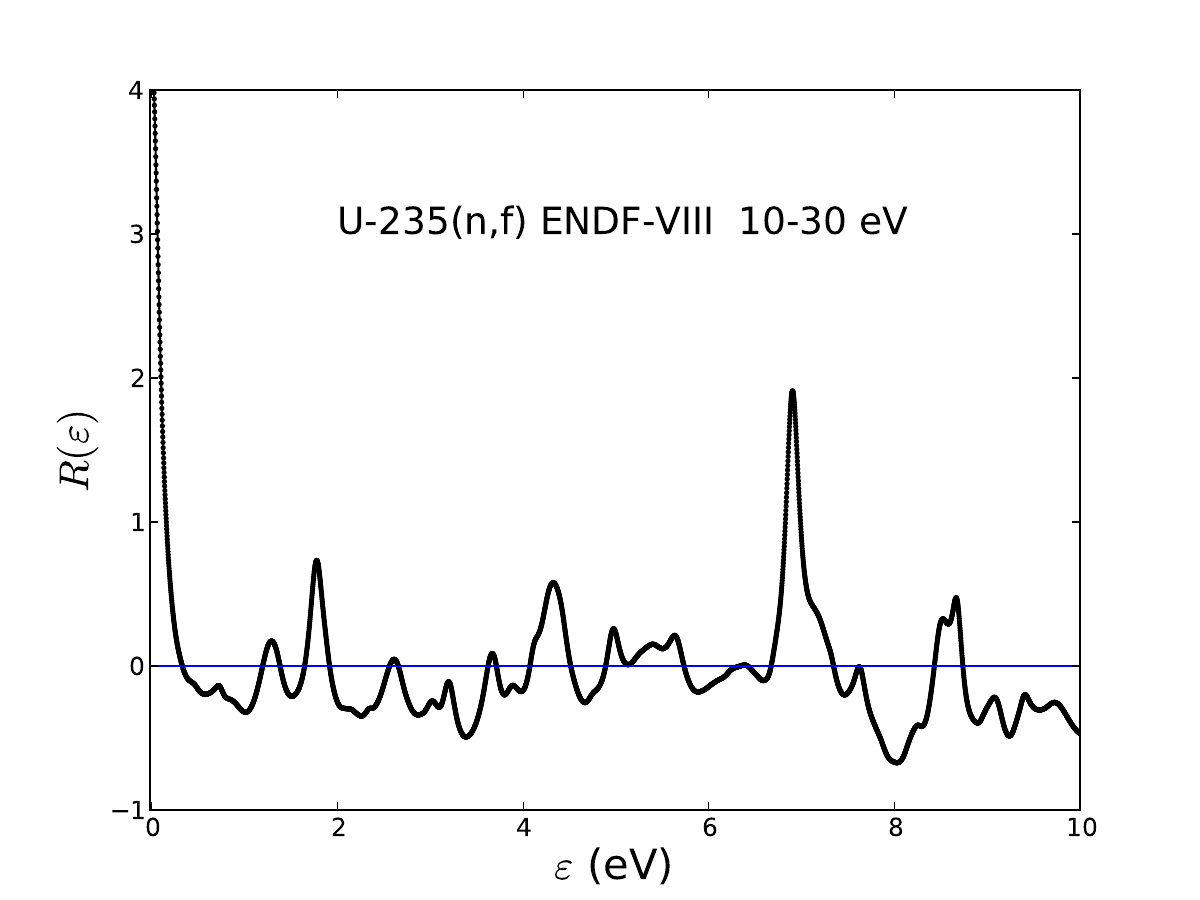}
\caption{Autocorrelation $R(\eps)$ for the data shown in Fig. 
\ref{sigma-endf} for the range $0 < \eps < B/2$.
}
\label{R-endf} 
\end{center} 
\end{figure}

As a warm-up to the computation of $R$ on multi-keV
energy intervals, we consider  fluctuations in the resolved
resonance region below 100 eV.  
Figure \ref{sigma-endf} shows the experimental cross section for the
neutron energy range 10--30 eV, with the data taken from the
ENDF-VIII.0 evaluated cross section \cite{ca18,br18,ca18b,Rmatrix}.  
The corresponding autocorrelation
function calculated from Eq. (\ref{R}) 
is shown in 
Fig. \ref{R-endf}.
There is a clear peak at $\eps = 0$.  Its shape parameters are $R(0) = 4.23$ 
and 
$\Gamma_{HM} = 0.09$ eV.  Table II gives the peak parameters
as well as those  for other energy
ranges below 100 eV.  
\begin{table}[htb] 
\begin{center} 
\begin{tabular}{|c|ccc|ccc|}
\hline
&  \multicolumn{3}{c|}{ENDF} & \multicolumn{3}{c|}{Mazama} \\
Energy range & $\sigma_0$(b)  & $R(0)$  &  $\Gamma_{HM}$(eV)  &
$\sigma_0$(b)  & $R(0)$  &  $\Gamma_{HM}$(eV)\\ 
\hline 
10 - 30 eV &  45.2  &  4.23     &  0.09  &$48\pm18$  &$3.6\pm 1.4 $  &
$0.10\pm0.02$ \\
30 -50  &  44.7  &  2.58     &  0.10   &&  & \\
50 -70  &  37.   &  2.40     &  0.16   && &  \\
70 -90  &  29.9  &  1.42     &  0.17   &$27\pm9$  &$3.2\pm 1.0$ & $0.10\pm0.02$  \\
\hline
\end{tabular} 
\caption{Autocorrelation peak height and width for various energy
ranges.
}
\label{peaks} 
\end{center} 
\end{table} 
Both $R(0)$ and $\Gamma_{HW}$ change with increasing energy $E$.  
There is also a prominent peak at $\eps = 7$ eV, which we interpret
as a statistical fluctuation of no physical significance.

To see how well one can understand the peak at $\eps = 0$,
we first compare with the prediction of the analytic
model, Eq.~(\ref{eq2}).  The reaction parameters for one entrance channel 
are given in Table \ref{params} \cite{footnote}.
\begin{table}[htb] 
\begin{center} 
\begin{tabular}{|c|cc|}
\hline
parameter & value & source \\ 
\hline 
$D=\bar D/2$               & 0.47 eV   & \cite{ca18} \\
$\bar \Gamma_\gamma$ & 0.038 eV & \cite{ca18} \\
$S_0$     & $0.98\times 10^{-4}$ eV$^{-1/2}$  & \cite{ca18} \\
$ \alpha^{-1}$ &  1.69 & \cite{nndc} \\
\hline 
\end{tabular} 
\caption{Reaction parameters for the neutron reactions on
$^{235}$U at low energy.  The parameter $\alpha^{-1} =
\langle\sigma(n,f)\rangle/\langle\sigma(n,\gamma)\rangle$ is evaluated
taking the averages over the energy range $E_n = \text{10--100}$ eV.
}
\label{params} 
\end{center} 
\end{table} 
The total width $\Gamma$ is the sum of the width in the entrance channel
$\Gamma_n$, the capture width $\Gamma_\gamma$, and the fission
width $\Gamma_f$.  The fission width be computed from the
parameters in Table II as $\Gamma_f \approx \alpha^{-1} \Gamma_\gamma
= 0.064$ eV.
The entrance-channel width is computed as
$\Gamma_n \approx S_{0} E^{-1/2} \pi/k_n^2$, where $k_n$ is the neutron
momentum in the entrance channel.  It is entirely negligible 
compared to $\Gamma_f + \Gamma_\gamma$ for
energies under  1 keV.  Thus, $\Gamma \approx \Gamma_\gamma
+ \Gamma_f = 0.102 $ eV over the region covering in Table I.
This yields $x = 0.94/(0.102\pi) = 2.9 $.  We also need the number
of channels and their weights 
to apply Eq.~\ref{eq:C0}.  There are two entrance
channels in the \unf~reaction at low energy, namely $J=3$ and
$J=4$, and it is reasonable to assume that their weights are close
to equal.  The situation is less clear for the fission channels.
Very likely there are only a few channels that contribute strongly.
Let us assume that there are three fission exit channels for each entrance channel,
and they all contribute equally.  Then the weighting factors in Eq. 
\ref{eq:C0} are  $\sum_{i,c} w_{ic}^2 = 1/6$ and 
$\sum_{i,c,c'} w_{ic} w_{ic'} = 1/2$.  Inserting these numbers in
Eq. \ref{eq:C0} we find  $R(0) \approx 7.0$.  This overestimate arises
because of our neglect of fluctuations in the total fission width,
which can be sizable if there are only three channels.  
These effects are included in the equations derived in Ref. \cite{er16},
but evaluating them requires several numerical integrations and
are not easy to condense to a simple formula.
Apart from that,
the observed variation of $R(0)$ over the 100 eV energy 
range of data set
is inexplicable.  None of the compound nucleus parameters vary
significantly on such a small energy scale, and the entrance
channels widths are small over the entire range.  

The correlation widths should be equal to the total widths of 
the compound nucleus resonances in the isolated resonance regime.
This appears to be the case for the first two energy ranges in
Table I, taking $\Gamma$ from the paragraph above.  However,
the extracted experimental width increases in the higher
energy ranges.  Again, this is not explicable given our
understanding of how the compound nucleus parameters vary with
energy.

One of us (GFB) has produced a Hamiltonian model 
that includes both a GOE treatment of the resonances and an
explicit Hamiltonian treatment of the entrance channel.
The code, called ``Mazama",  is briefly described in the
Appendix together with the Hamiltonian parameters as applied
to this work.  All the fluctuations inherent in the GOE 
 are included, since the code operates by sampling
the ensemble and calculating the full energy-dependent
cross section for each representative Hamiltonian.
The parameters of the ensemble are tuned to 
reproduce
the experimental $D$, $\Gamma_\gamma$ and $\alpha^{-1}$.
\begin{figure}[tb] 
\begin{center} 
\includegraphics[width=0.5\columnwidth]{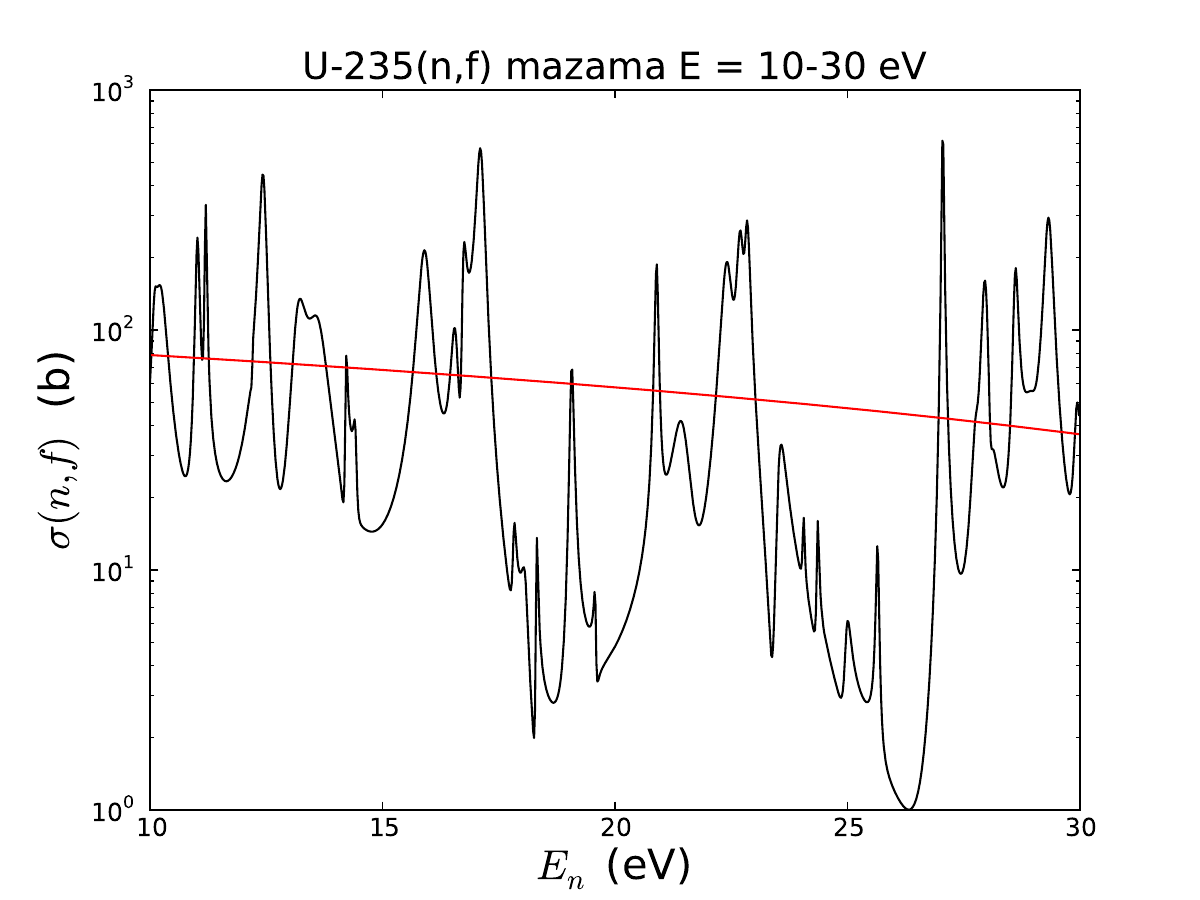}
\caption{Sample $^{235}$U(n,f) cross section calculated by the mazama
code.  The results are the average for two entrance channels.  Code
parameters are given in the Appendix.}
\label{sigma-mazama} 
\end{center} 
\end{figure} 
\begin{figure}[tb] 
\begin{center} 
\includegraphics[width=0.5\columnwidth]{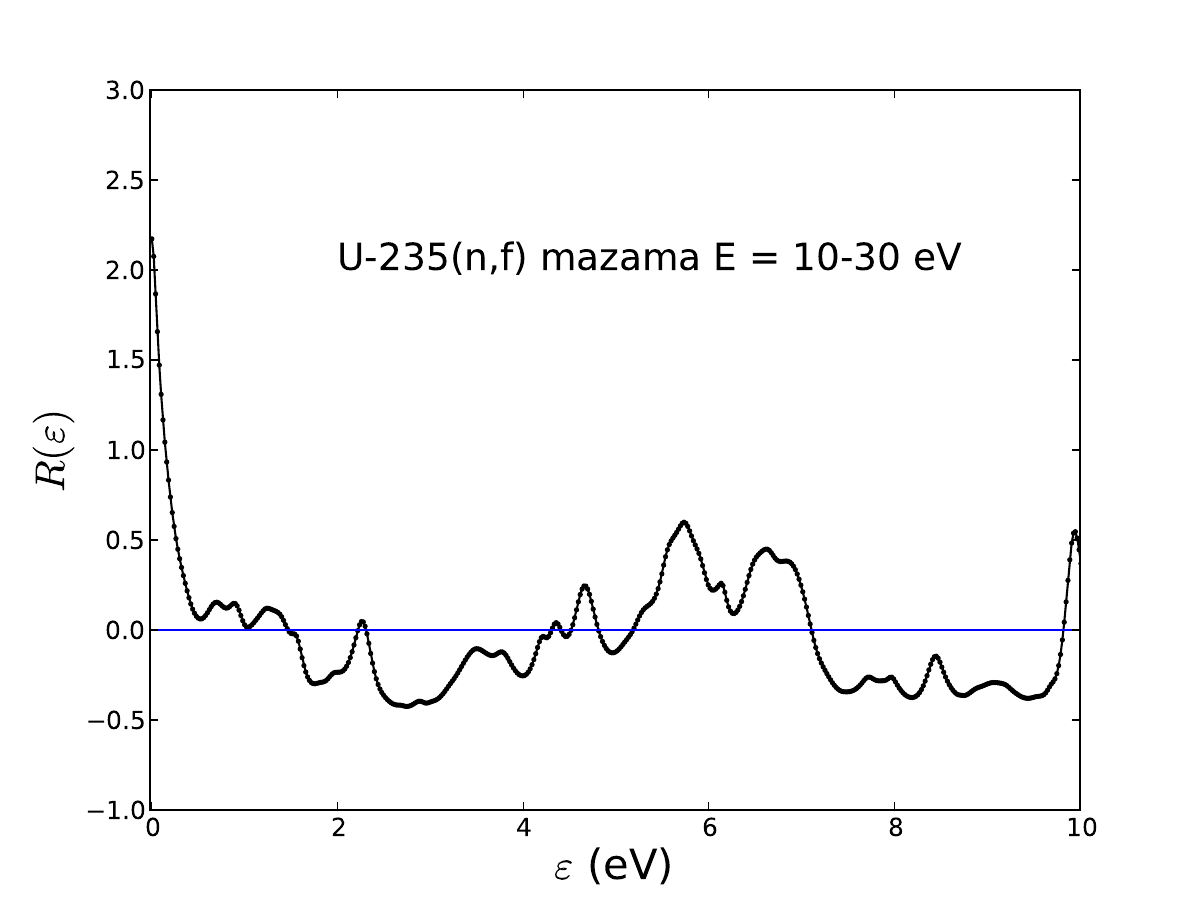}
\caption{
Autocorrelation $R(\eps)$ for the data shown in Fig. \ref{sigma-mazama}. 
}
\label{R-mazama} 
\end{center} 
\end{figure} 
The number of fission channels is taken to be 3.
A typical cross section produced by Mazama is shown in Fig.
\ref{sigma-mazama}.
The calculation was done by taking two samples from the ensemble, 
for the two independent entrance channels.
The extracted autocorrelation function is shown in 
Fig. \ref{R-mazama}.  
The related peak parameters are $R(0) = 3.6$ and $\Gamma_{HM} = 0.10$,
in fair agreement with experimental data at the lowest energy.  There are
of course uncertainties in the predicted peak parameters due to 
the Porter-Thomas fluctuations.  These can be estimated 
by taking many samples from the
GOE ensemble, evaluating the parameters for each sample, and then 
finding the rms dispersion in them.
We have
carried out this analysis for 100 samples to determine a 
mean and dispersion  of the parameters.  The results are
given in the leftmost columns of Table I.  The error bars on the extracted quantities
are the standard deviations of the individual samples.
The ensemble averages for the 10--30 eV interval agree with the experimental
numbers for the cross section and $R(0)$ value to within the error
bars. The Mazama analysis confirms that the GOE model has the potential
to describe data on $R(0)$.

The GOE model also shows that the uncertainty in the extracted
$R(0)$ can be uncomfortably large.  
The variance in the calculated  $R(0)$ is in the 30--40 \% range 
with about $40$ resonances in the energy window $B= 20 $ eV.
Assuming that standard statistics applies for errors,  we
estimate that the window should include about 500 resonances to reduce
the uncertainty in $R(0)$ to the 10\% level.    

One can also see from Mazama model
results for the 70--90 eV interval that the theoretical quantities 
hardly vary at all from those of the 10--30 eV interval.  This is
not unexpected, since the only real difference is the increased 
penetrability in the entrance channel.  This substantiates our earlier
assertion that the strong energy dependence of the $R(0)$ and 
$\Gamma_{HM}$ obtained from ENDF cross sections is completely
inexplicable in standard models of compound nucleus reactions. However, 
for the purpose of our subsequent discussion of cross section fluctuations at multi-keV energies, we shall
take the point of view that, as regards the 10--30 eV data and its modeling
by the Mazama code, a good  account is given of the primary peak in $R$. It is unambiguously associated with
fluctuations in properties of individual compound-nucleus
resonances.

\section{Higher energy region}

We now go to the multi-keV energy region where evidence for
fluctuations on the keV energy scale were reported.  We focus on 
the experimental data in Ref.~\cite{pe74} which covers the
energy range 2--100 keV.  We first examine the data
in the energy interval 10--25 keV that was shown in Fig. 1. 

\begin{figure}[tb] 
\begin{center} 
\includegraphics[width=0.5\columnwidth]{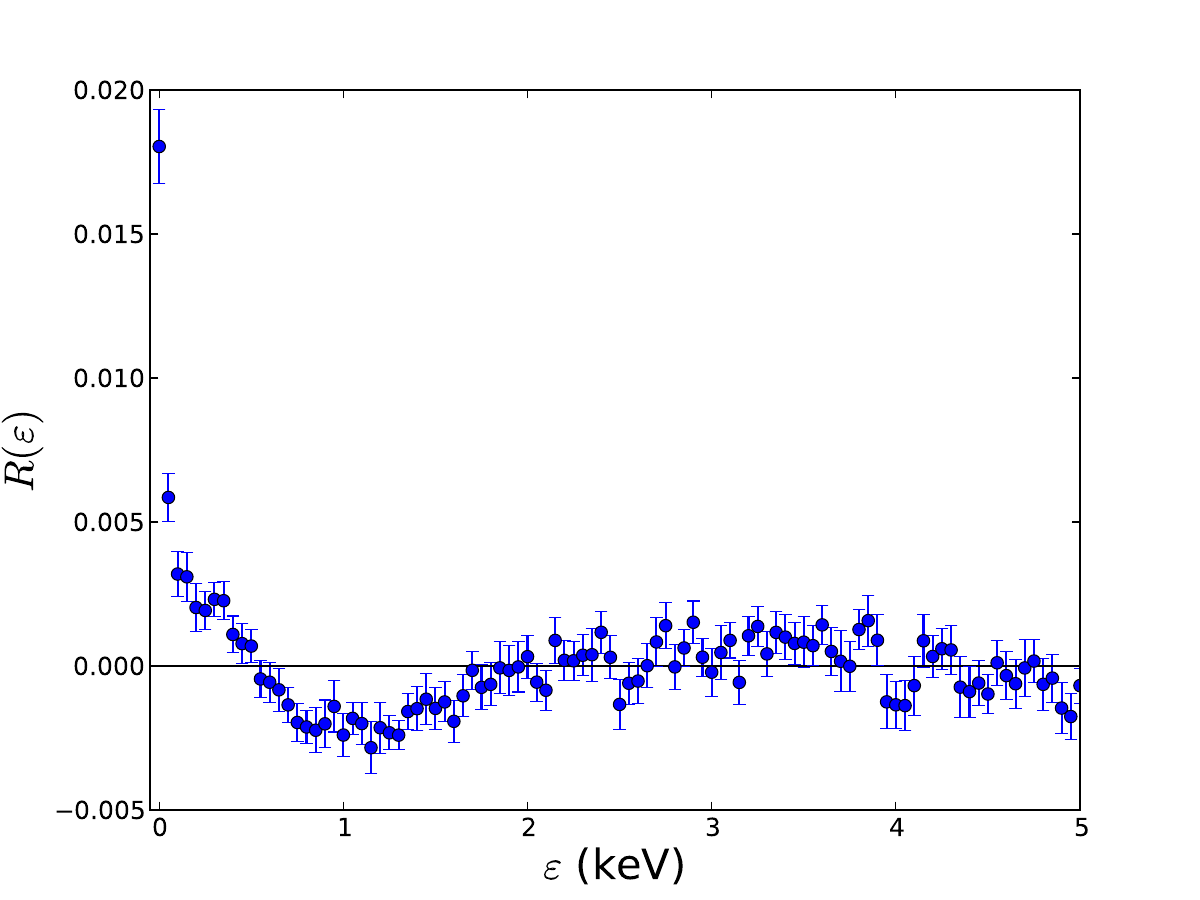}
\caption{Autocorrelation function for the \unf~reaction over
the energy range $10 < E < 25$ keV, using the cross section
data of Ref. \cite{pe74} as given in Ref. \cite{nndc}.  
}
\label{R-10-25keV} 
\end{center} 
\end{figure}

It is safe to assume that the resonance spacing $D$
in this higher-energy window is unchanged from its value 
at energies below 100 eV. Then the number of resonances
is about 30,000, giving adequate statistics for measuring
$R(0)$.  The bin size is $\Delta E = 0.05$ keV, and the
autocorrelation function is computed with the same binning.
For presenting the derived $R$, we take
an ensemble of cross section data
sets generated by adding a random error to the measured
ones, taking its variance 
from the tabulated experimental data.  The results are
shown in Fig.~\ref{R-10-25keV}.  
\begin{figure}[tb] 
\begin{center} 
\includegraphics[width=0.5\columnwidth]{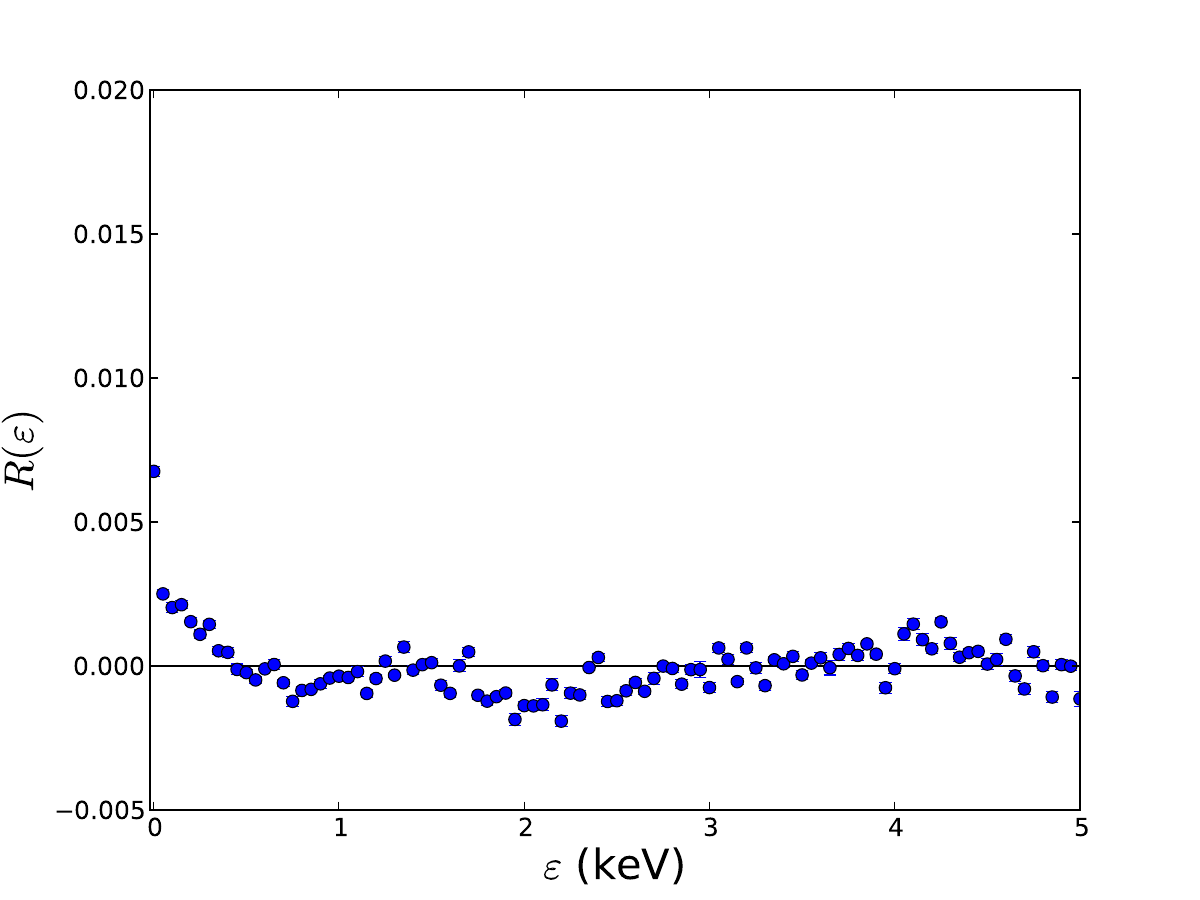}
\caption{Same as Fig. \ref{R-10-25keV} using the data from
Ref. \cite{bo71}. The error here is as small or smaller than
the plotted points.
}
\label{R-10-25keVb} 
\end{center} 
\end{figure} 
There is a clear peaking near $\eps = 0$ that is much
larger than the experimental error bars. 
Thus, the measurement provides quantitative  
information about the correlation function that can hopefully
can used to gain a better understanding of the reaction
dynamics.  

The peak in Fig.~\ref{R-10-25keV} has a double structure, namely, a
single-bin spike at $R(0)$, and then a broader ``shoulder" 
extending out to about 0.5 keV.  The height of the $R(0)$ spike
is about 0.02.  Equation (\ref{eqRDE})  should be applicable
for
the contribution of the compound-nuclear resonances.  With
the given $D$ and $\Delta E$, Eq.~(\ref{eqRDE})  gives 0.02, in agreement
with the data.  An unexpected feature of $R(\eps)$ in Fig.~\ref{R-10-25keV}
is the dip below zero between $\eps\approx 0.5\,$keV and
 $\eps\approx 1.5\,$keV. In fact, for certain parameter choices,
Eq (\ref{eq2})  does admit a negative dip in $R$ when the
level repulsion terms are dominant ($\eps\sim D$).  However, as $D$ sets
the location for any such dip, it would not be visible on the
keV energy scale of Fig.~\ref{R-10-25keV}.  

Next we examine the data of Ref. \cite{bo71}, which also
has enough energy resolution 
to resolve the peak structure.  Its autocorrelation function
is shown in Fig. \ref{R-10-25keVb}.
It confirms in a qualitative way the $R(0)$ spike and the broader 
structure out to $0.5$ keV.  However, it shows no 
dip distinguishable from the excursions from
zero at larger $\eps$.  In this data set, the height of 
the $R(0)$ spike before the broader structure is 0.005, a factor
of 4 smaller than the $R(0)$ spike for the previous data.  It is difficult to
reconcile the disagreement here.  In any case, the presence
of the broader shoulder structure is confirmed, showing that there
is physics present beyond the statistics of the compound nucleus.

\begin{figure}[tb] 
\begin{center} 
\includegraphics[width=0.5\columnwidth]{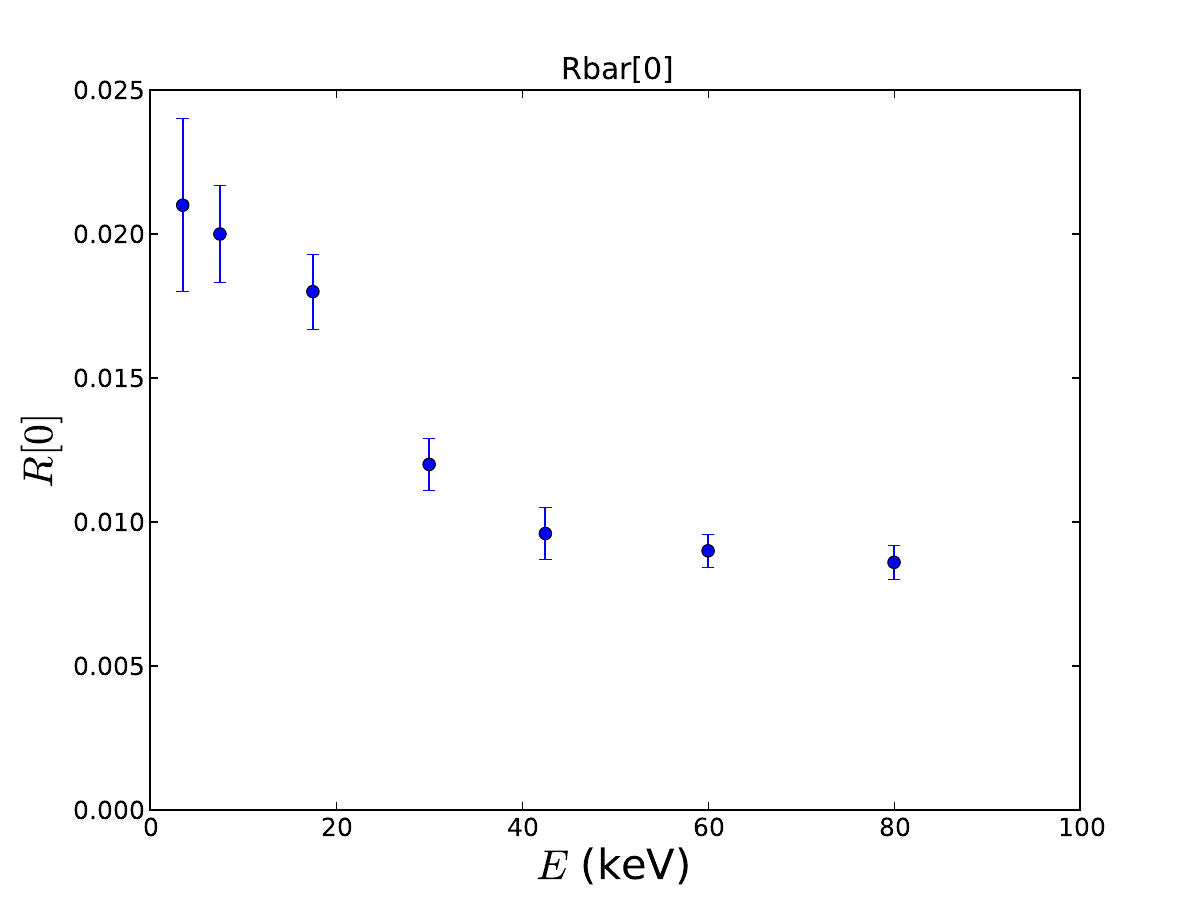} 
\caption{Autocorrelation spike for the data of Ref.
\cite{pe74}, showing several energy windows, beginning with 2--5 keV 
and ending with 70--90 keV.  Error bars are computed from
the quoted experimental errors in the data as described
in the text.}
\label{Rbar0} 
\end{center} 
\end{figure}

\begin{figure}[tb] 
\begin{center} 
\includegraphics[width=0.5\columnwidth]{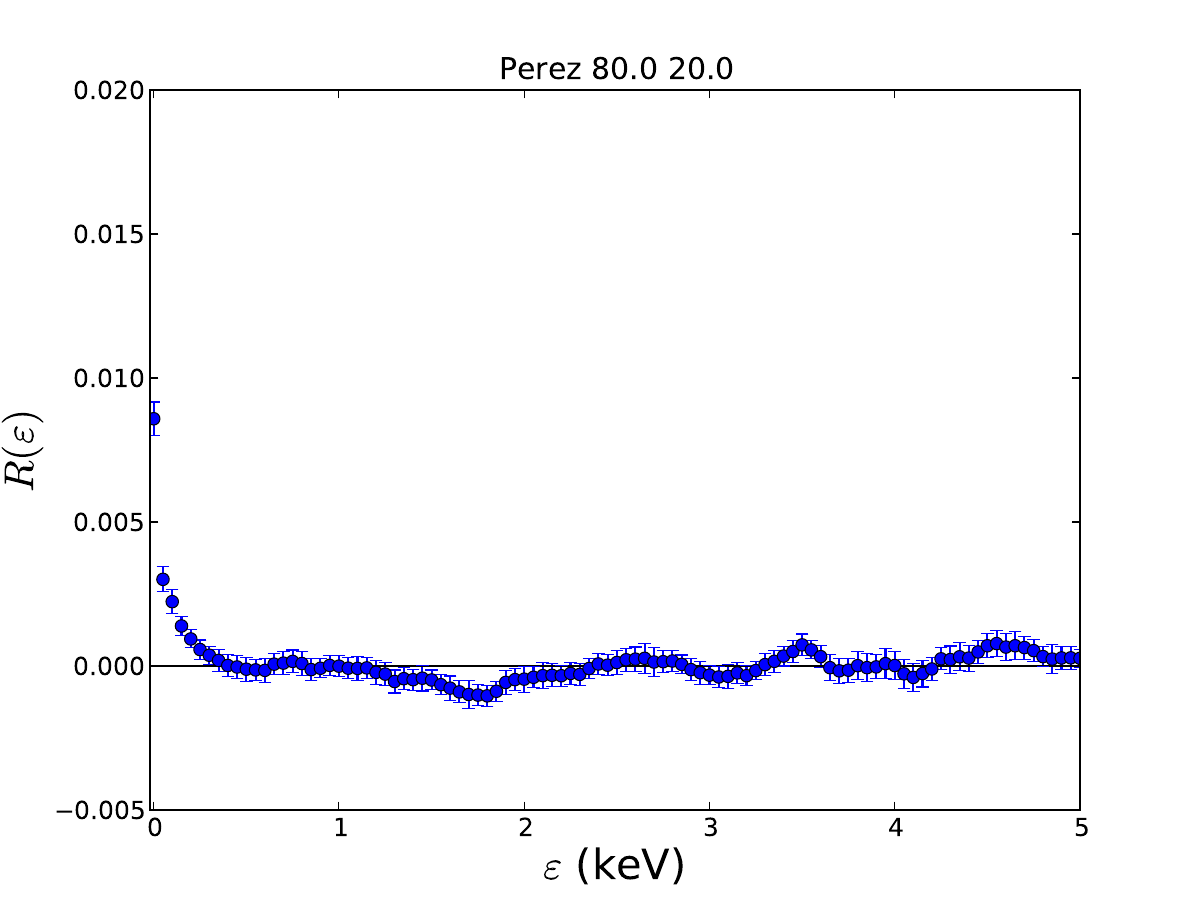}
\caption{Same as Fig. \ref{R-10-25keV} for an energy window
extending from 70 keV to 90 keV.
}
\label{R-70-90keV} 
\end{center} 
\end{figure}

The data of Ref. \cite{pe74} extends over the entire energy span of interest to us (from 2 to 100
keV). For the purposes of analysis, we have subdivided this energy interval
into seven windows ranging from 3 to 20 
keV in width.  Only the 10--25 keV window that shows a distinctive peak beyond
the $R(0)$ spike.  Thus, if the
structure in that region is due to doorway resonances (beyond
the compound nucleus reaction mechanism), then the doorway state spacing
must be wider than tens of keV. The
$R(0)$ spike persists in all the windows.  This may be seen
in Fig.~\ref{Rbar0}, showing the spike height for each of the seven
windows we examined.  Up to the 10--25 keV window already 
discussed, the spike height agrees with the expected 
compound nucleus cross section fluctuations.  At higher
energies, the peak decreases.  Part of the decrease is 
undoubtedly a degraded energy resolution that mixes strength 
into adjacent energy bins. This is particularly apparent in
the highest window we examined, shown in Fig. \ref{R-70-90keV}.
Here the window extends from 70 to 90 keV and the energy
bins have a width $\delta E = 0.05$ keV.   
The compound-nucleus peak could also decrease if more 
entrance channels participate in the reaction.  At some
point $p$-wave neutron capture will become significant,
but we haven't investigated whether it will affect
the autocorrelation on the 100 keV energy scale. 

\section{Conclusion and outlook}

While we confirm the effect of compound-resonances on the
autocorrelation function up to near 100 keV in neutron
bombarding energy, we see no evidence for correlations
on the scale of 1 keV, contrary to the claim of Ref.
\cite{pe74}.  Indeed, the structure we see at all is
isolated in the 10--25 keV energy window.  

Neutron-induced fission is very different below the
barrier.  There states in the second well mediate
barrier penetration producing broad resonances in the
fission cross sections.  It might be that these resonances
persist above the barrier in an attenuated form.
That could be a plausible explanation of the 
22 keV structure, provided the resonance spacing in
the second well is compatible with the non-observation
of other structure in the 2--100 keV energy range.  Perhaps
more experimental study of fission cross sections
around the barrier top would clarify the situation.  
In that respect photofission is
a good tool, because it can reach excitation energies below
those accessible by neutron-induced fission.

\section*{Acknowledgements} 
Work at Brookhaven National Laboratory was sponsored by the Office of
Nuclear Physics, Office of Science of the U.S. Department of Energy under
Contract No. DE-AC02- 98CH10886 with Brookhaven Science Associates, LLC.
E.D.D.~was supported in part  by the U.S.~Department of Energy under Grant 
No.~DE-FG02-97ER41042.

\section*{Appendix} 

\subsection{Overview of experiments}

In this work we focused on the three highest resolution absolute $^{235}$U(n,f) cross section data sets available in the
 range 1--100 keV~\cite{bo71,mo78,pe74}:
\begin{itemize}
    \item The measurement in Ref.~\cite{bo71} was performed at the Lawrence Radiation Laboratory (now known as LLNL) using the Livermore 100 MeV electron linac and a 200 m flight path.
    \item The measurements in Refs.~\cite{pe74} and \cite{mo78} were performed at the ORELA facility at ORNL.  Ref.
    \cite{pe74} reported a conventional (n,f) measurement using a 150 m flight path. The experiment of Ref.
    \cite{mo78} was unconventional, using a polarized neutron beam 
impinging on a polarized target. This permitted the determination of the spins of individual resonances. A much shorter flight path of approximately 14 m was used in Ref.~\cite{mo78}, but the decrease in energy resolution from the reduced flight path was more than compensated by cryogenically cooling the polarized target to 0.1 K.
\end{itemize}

In all three cases, the neutrons were produced by photonuclear reactions
induced by incident electron bremsstrahlung emission and then passed through
a light water moderator.  In each experiment, the targets were enriched in
$^{235}$U.  Contaminants (either $^{238}$U or $^{16}$O for the oxide
targets) were properly accounted for.

There are other experiments reporting data in the energy range of interest to us, but we found
them less informative for a variety of reasons:
\begin{itemize}
    \item The experiments of Refs. \cite{Blons,Mostovaya,Weston,Gayther,Patrick} all have 
    inferior energy resolution.  The data in Ref.~\cite{Gayther} were actually taken as a scoping study for the experiment in Ref.~\cite{Patrick}.
    There are other experiments in this energy range that have substantially 
    worse  energy resolution and are not listed in Table \ref{U235ExpSummary}.
    \item Ref.~\cite{Brown} lacks adequate published documentation.
    \item The uncertainty data needed for evaluation purposes was either missing
    or not understandable from the experiments reported in 
    Refs.~\cite{Migneco,Wagemans,Wang,Bowman1970}.
   The data in Ref.~\cite{Bowman1970} were actually taken as a scoping study for the experiment in Ref.~\cite{bo71} used in our analysis.
    \item Ref.~\cite{Gwin} was withdrawn.
    \item Refs.~\cite{Albert,Lemley,Cramer,Brown} used a nuclear explosive as the neutron source and therefore have unquantifiable uncertainties in both the flight path used and the neutron fluence.
\end{itemize}
Table \ref{U235ExpSummary} summarizes these experiments.  As is usual for neutron resonance measurements, the incident neutron energy in all cases was determined using time of flight (ToF).

Fr\"ohner and Haddad \cite{Froehner} performed a detailed study of sources of uncertainty in ToF measurements.  They argue that the uncertainty on the determined incident energy is given by $\Delta E/E = \sqrt{a+bE}$ with constants $a$ and $b$.  Here, $a$ depends on overall flight path length, the neutron production target thickness and the moderator thickness surrounding the neutron production target.  The constant $b$ depends on the rescattering time in the neutron source, moderator and target (including thermally induced jitter) and the overall flight path length.  In all of the experiments considered, these effects were carefully considered and we believe the reported energy resolutions for the experiments are reasonable.

Leal \emph{et al.}~\cite{Leal} additionally advocate a cross section normalization factor of the form $a+b\sqrt{E}$.  Given the limited energy range over which we considered cross section data, the additional $\sqrt{E}$ dependence was not found to be needed.
\begin{table}
	\scriptsize
    \begin{tabular}{lp{0.3in}p{1.2in}p{0.6in}p{1.5in}p{0.7in}p{0.7in}}
        \hline\hline
        Ref. 		  & Pub. Year 		& Author                        & EXFOR (sub)Entry  & Facility/Laboratory            & Flight Path   & Energy resolution (keV)\\
        \hline\hline
        \cite{bo71} & 1971&	           C. D. Bowman \emph{et al.}      		& 10419.002	        &LRL (now LLNL) Electron Linac	 & 250 m	            & 0.010-0.007\\
        \cite{pe74}  & 1974&	           R. B. Perez \emph{et al.}   			& 10302.002	        &ORELA, ORNL	                 & $151.9\pm 0.1$ m	 	& 0.025\\
        \cite{mo78}  & 1978&	           M. S. Moore \emph{et al.}       		& 10629.004	        &ORELA, ORNL	 & \text{13.40 m}\tablenote{\text{Fssion monitor}}, 15.28 m\tablenote{\text{Transmission monitor}} 	& 0.00005\\
        \hline
        \cite{Wang}   & 1965&              Wang \emph{et al.}                  & 40271.003         &JINR, Dubna                     & 1000 m              	& 0.05 \\
        \cite{Albert} & 1966&	         R. D. Albert                  & 12343.002	        &AGT, LRL (now LLNL)	         & 1280 km	          	& 0.10\\
        \cite{Brown}  & 1966&	           W. K. Brown \emph{et al.}      		& 12432.002	        &UGT, LASL (now LANL)	         & Unknown	          	& 0.033\\
        \cite{Patrick} & 1970 &            B. H. Patrick \emph{et al.} 			& 20461.002 		&Harwell 45 MeV Linac 			 & 97.5 m & 0.18  \\
        \cite{Cramer} & 1970 & 		       J. D. Cramer          			& 10057.004 		&UGT, LASL (now LANL) 			 & 214.43 m & 0.1  \\ 
	    \cite{Bowman1970} & 1970 & 		   C. D. Bowman \emph{et al.}  			& 10170.002, .003   &LRL (now LLNL) Electron Linac	 & 250 m				& 0.05-0.1 \\
        \cite{Lemley} & 1971&	           J. R. Lemley \emph{et al.}   		& 10120.002	        &UGT, LASL (now LANL)	         & Unclear\tablenote{\text{Only distance to borehole given}}		& 0.004\\
        \cite{Blons} & 1971&	           J. Blons \emph{et al.}         		& 20483.002	        &60 MeV Saclay LINAC	         & 50.07 m	          	& ${\sim 0.9}$\\
        \cite{Gayther} & 1972 &			   D. B. Gayther \emph{et al.}      	& 20422.002 		&Harwell 45 MeV Linac 			 & 97.5 m 				& 0.25  \\
        \cite{Migneco} & 1975&	           E. Migneco \emph{et al.}     		& 20783.002	        &CBNM Linear Accelerator, Geel	 & 60.58 m	          	& 0.0007\\
        \cite{Gwin} & 1976&	               R. Gwin \emph{et al.}         		& 10267.024	        &ORELA, ORNL	                 & 40 m	              	& 0.009\\
        \cite{Wagemans} & 1976&            \hbox{C. Wagemans,} \hbox{A. J. Deruytter}	& 20826.004         &CBNM Linear Accelerator, Geel   & 30 m                	& 0.025 \\[-1pc]
        \cite{Mostovaya}& 1980&	           T. A. Mostovaya \emph{et al.}		& 40616.004	        &Electron Linear Accelerator `FAKEL', Kiev	&26 cm	  	& 0.06\\
        \cite{Weston} & 1984&	           \hbox{L. W. Weston,} \hbox{J. H. Todd}         & 12877.004	        &ORELA, ORNL	                 & 20 m	              	& 0.10\\[-1pc]
        \hline\hline
    \end{tabular}
    \caption[]{\label{U235ExpSummary} 
Absolute  $^{235}$U(n,f) cross section measurements in the range 10--25 keV,
retrieved from the EXFOR library \cite{EXFOR}.  Only the first three
measurements, above the line in the table, are used in the present analysis. 
See text for details.}
\end{table}

\subsection{Simplified formula for $C(0)$}\label{simple_formula}

\begin{figure}[t!] 
\begin{center} 
\includegraphics[width=0.5\columnwidth]{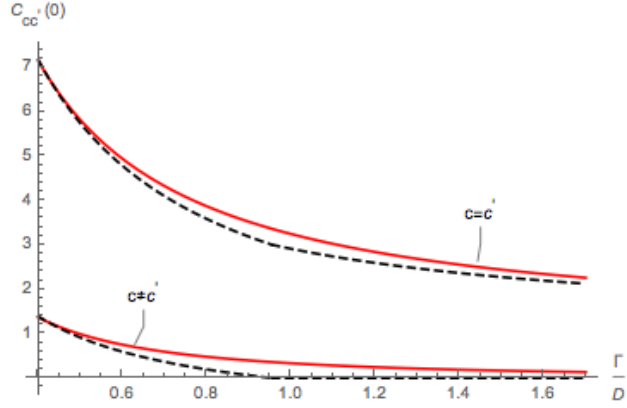}
\caption{Plot of $C_{icc^{\prime}}(0)$  versus $\Gamma/\bar D\ (\ge 0.4)$.  Solid lines: evaluation of Eq.~(\ref{eq2}). 
Dashed lines: approximation in Eq.~(\ref{eq:Cga}).
}
\label{fg:C0anatomy} 
\end{center} 
\end{figure}

\begin{figure}[b!] 
\begin{center} 
\includegraphics[width=0.5\columnwidth]{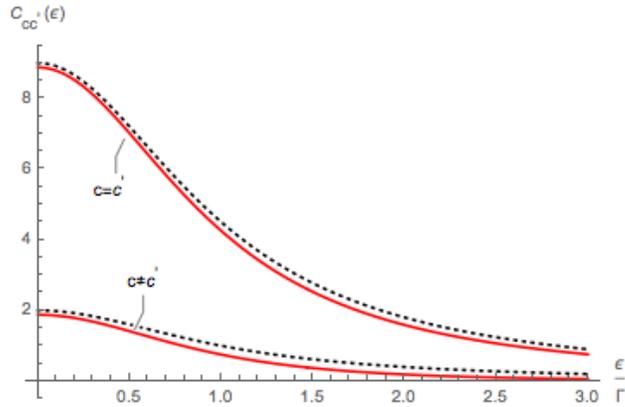}
\caption{Plot of $C_{icc^{\prime}}(\eps)$  versus $\varepsilon/\Gamma$ for $x=1$ (a value of $x$ appropriate to the
resolved resonance regime of ${}^{235}$U). Solid lines: evaluation of Eq.~(\ref{eq2}). 
Dotted lines: approximation in Eq.~(\ref{eq:Cga}).
}
\label{anatomy} 
\end{center} 
\end{figure}

Equation (\ref{simple_f}) was obtained by stitching together the two expressions
valid in the limits  $x \gg 1$ and $x \ll 1$. 
The effectiveness of this approximation 
is illustrated by the plot of $C_{icc^{\prime}}(0)$ versus  $\Gamma/\bar D$ in Fig.~\ref{fg:C0anatomy}. 
The approximation is poorest when the magnitude of $\Gamma$ approaches that of
$\bar D$, as is to be expected. 
Figure~\ref{anatomy} 
contains complementary information on the extent to which the $\eps$-dependence of 
$C_{icc^{\prime}}(\eps)$ is reproduced when the approximation to $C_{icc^{\prime}}(0)$ is good.
 
Errors in the approximation  of $C_{ic{\not=}c^{\prime}}(\eps)$ by Eq.~(\ref{eq:Cga}) can be sizeable 
($\sim 100\%$), but only when it is an order of magnitude smaller than $C_{icc}(\eps)$. The approximation of
$C_{icc}(\eps)$ is, by contrast, always reasonable, the percentage error being never more than 10\% 
(even when $\Gamma\simeq \bar D$).  For the reaction studied here, there are
only a few fission channels and $C_{icc}$ term should dominate.
Thus, Eq.~(\ref{eq:Cga}) should
suffice to approximate a calculation of $C(\eps)$ based on Eq.~(\ref{eq2}) to within an error of 10\% or so.

\subsection{Effect of finite energy resolution }\label{enres}
 
\begin{figure}[b!] 
\begin{center} 
\includegraphics[width=0.5\columnwidth]{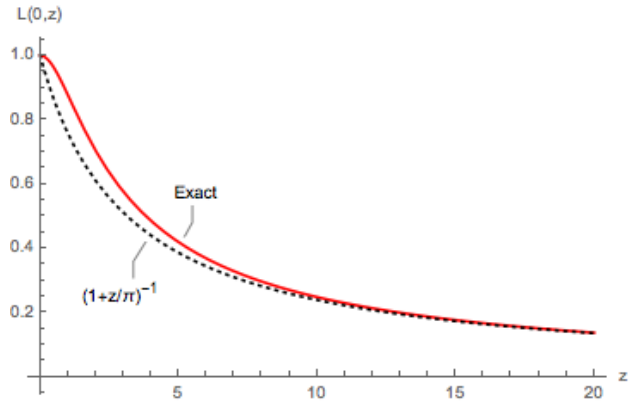}
\caption{Plot of $L(0,z)=[2z\arctan(z)-\ln(1+z^2)]/z^2$  versus $z$. Solid line: exact evaluation of $L(0,z)$. 
Dotted line: approximation in Eq.~(\ref{eq:L0zAPP}).
}
\label{fg:L0z} 
\end{center} 
\end{figure} 

Under the assumption that cross sections are reported
as averages over energy bins of fixed width $\Delta E$,
 $R(\eps)$ in Eq.~(\ref{eq:Csummary}) 
is replaced by
\be
 R(\eps,\Delta E) = R(0)\,  L\!\left(\tfrac{\eps}{\Gamma},\tfrac{\Delta E}{\Gamma}\right),
\ee
where~\cite{Gi65}
\be
 \begin{split}
   L(y,z) =\, &\frac{1}{z^2}\, \Biggl[ (y+z)\arctan(y+z)+(y-z)\arctan(y-z)- 2y \arctan(y)\\ 
               & \qquad -\tfrac{1}{2}\ln\frac{1+(y+z)^2}{1+y^2} -\tfrac{1}{2}\ln\frac{1+(y-z)^2}{1+y^2} \Biggr] .
\end{split}
\ee
With a view to establishing the properties of $L(y,z)$,
it is helpful to study $L(0,z)$ and $l(y,z)=L(y,z)/L(0,z)$ separately. As noted by Gibbs~\cite{Gi65},
a simple approximation of $L(0,z)$ is viable (see Fig.~\ref{fg:L0z}), namely,
\be \label{eq:L0zAPP}
  L(0,z)\approx \frac{1}{1+ z/\pi} .
\ee
The qualitative character of $l(y,z)$ depends on the magnitude of $z$. It is the
behavior for small and large $z$, which is of interest to us. For $z<\tfrac{1}{2}$, 
$l(y,z)\approx (1+y^2)^{-1}$, i.e., as one would expect for good energy resolution,
the dependence on $\eps$ is essentially unchanged from that in Eq.~(\ref{eq:Csummary}).
For $z\gg 1$, the variable $\rho=y/z$
is preferred to $y$ (appropriate for $z\lesssim 1$); in terms of physical parameters, 
this amounts to relating $R(\eps,\Delta E)$ to $\eps/\Delta E$ and $\Delta E/\Gamma$, not
$\eps/\Gamma$ and $\Delta E/\Gamma$. 

\begin{figure}[b!] 
\begin{center} 
\includegraphics[width=0.5\columnwidth]{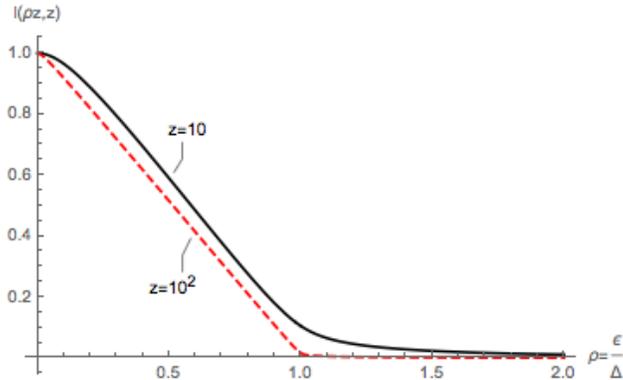}
\caption{Plot of $l(\rho z,z)$  versus $\rho=\eps/\Delta E$ for two choices of $z=\Delta E/\Gamma\ (\gg 1)$.
}
\label{fg:Linf} 
\end{center} 
\end{figure} 

The limit 
\be   \label{eq:limC}
  \lim\limits_{z\rightarrow\infty} l(\rho z, z) = (1-|\rho|) \Theta(1-|\rho|)\qquad (\rho\  \text{fixed})
\ee
forms the basis for an approximate representation of $R(\eps,\Delta E)$. Figure \ref{fg:Linf}
demonstrates how rapidly this limit is attained. Combining our results on $L(y,z)$,
we obtain Eq.~(\ref{eqRDE}).

\subsection{Realistic Modeling}

Here we provide details of the Mazama model used to
compute the cross section shown in Fig. \ref{sigma-mazama} and the
autocorrelation parameters listed in Table \ref{peaks}.  The model
is defined by a matrix Hamiltonian acting in a space
comprising the entrance channel wave function on a 
coordinate-space mesh $r_i$ and one or more sets of
internal states.  Compound nucleus states are identified
with GOE Hamiltonian eigenstates.  
To the real GOE resonance
energies are added fixed imaginary energies associated
with the gamma decay widths.  In addition, the Hamiltonian
includes coupling to one or more fission channels, represented as
discrete doorway states that couple to the compound nucleus states.
The inelastic $S$-matrix elements are found from solving
a Schr\"odinger equation with a boundary condition on the
entrance-channel wave function.

The key physical parameters
are:
\begin{itemize}
\item the Woods-Saxon potential in the entrance channel sector
characterized by the usual parameters ($V_{ws},a_0,R_0 = r_0 A^{1/3}$);
\item the average level spacing $\bar D$ of the compound-nucleus states populated from an
  entrance channel;
\item the gamma decay width of the compound-nucleus states, assumed to be the same for all states;
\item the number of the fission doorway states (each representing a fission channel) and their fission decay widths;
\item the Porter-Thomas distributed coupling matrix elements between the entrance
channel and the compound-nucleus states;
\item the Porter-Thomas distributed coupling matrix elements between the compound-nucleus states and the fission doorway states.
\end{itemize}
For the calculations reported here, the Woods-Saxon parameters are close to those obtained
by a global fit of single-particle properties at the Fermi surface \cite{BM}.  The
level spacing $D$ and $\Gamma_\gamma$ are the same as in Table I.  The
coupling matrix elements between the compound-nucleus states $|\mu\rangle$
and the
entrance channel $|n\rangle$ are parameterized as 
\be
\langle n,i | v \ \mu\rangle  = v_n s_\mu ,
\ee
where
$i$ is a mesh point close to the nuclear
surface and $s_\mu$ is a Gaussian random variable of unit variance.
The strength $v_n$ is fitted to the integrated inelastic cross section over the
energy interval 10--100 eV.  

There is considerable ambiguity in choosing the parameters associated with
the fission channels.  In this work we assume that there are 3 fission
channels, each coupled to the compound nucleus states with matrix element
$\langle \mu | v | f\rangle =  v_f s_\mu$. The strength $v_f$ 
is chosen to make the average mixing between the fission channels and the
compound states uniform. In effect, the fission channels are part of the 
GOE, but with different
decay widths.  This leaves a single parameter to be determined, namely
the decay width of the fission channels.  We determine it
by fitting to $\alpha^{-1}$ from Table I.  The values of 
the parameters 
are given in Table \ref{params}.
\begin{table}[htb] 
\begin{center} 
\begin{tabular}{|c|c|}
\hline
parameter & value  \\ 
\hline 
resonance spacing & $D = 0.94$ eV    \\
capture width  & $\bar \Gamma_\gamma = 0.038$ eV  \\
Woods-Saxon well & $V_0  = 44$ MeV  \\
"&  $ a_0 = 0.65 $ fm   \\
" &  $ r_0 = 1.25 $ fm  \\
n-c coupling & $v_n = 2.2$ keV  \\
c-f coupling & $v_f = 5.1$ eV  \\
fission width, per channel & $ \bar \Gamma_f = 0.040$ eV \\
no. of fission channels &  $N_f = 3$ \\ 
mesh spacing & $\Delta r =   0.5$ fm     \\
\hline 
\end{tabular} 
\caption{Hamiltonian parameters for simulating neutron reactions on 
$^{235}$U with the Mazama code.  All parameters except $v_n$ are insensitive
to the mesh spacing of the entrance channel.
}
\label{paramsj} 
\end{center} 
\end{table}


\begin{thebibliography}{99}
\bibitem{footnote}  Note that $D = 0.47$ includes resonances from
both entrance channels  ($J^\pi = 3^-$ and $4^-$).
\bibitem{Rmatrix}  The ENDF evaluation also has individual resonance
parameters, but we also use the reconstructed cross section 
to extract the autocorrelation function.
\bibitem{Typo} A plus sign should precede the ``1'' at the end of the second line of Eq.~(B12) in Ref.~\cite{er16}.
\end{thebibliography}
\end{document}